\documentclass[journal]{IEEEtran}
\usepackage{latexsym,,amssymb,amsmath,graphicx,epsf,cite,bbm,float}
\usepackage{ifpdf}
\usepackage{epstopdf,mathtools}

\usepackage{algorithm,algorithmic}
\usepackage{amsmath,amssymb,bm}
\usepackage{amsfonts,dsfont,color,bbm,subcaption}

\def\ninept{\def\baselinestretch{1}}
\ninept

\newcommand{\abs}[1]{|#1|}

\DeclareMathOperator*{\argmax}{arg\,max}
\DeclareMathOperator*{\argmin}{arg\,min}

\usepackage{hyperref,amsthm}

\newtheorem{theorem}{Theorem}

\newtheorem{lemma}[]{Lemma}

\newtheorem{proposition}[]{Proposition}

\newtheorem{corollary}[]{Corollary}

\newtheorem{remark}[]{Remark}

\newtheorem{definition}[]{Definition}

\newtheorem{assumption}[]{Assumption}

\newtheorem{notion}[]{Notion}

\begin{document}

\title{Natural Hierarchical Cluster Analysis by Nearest Neighbors with Near-Linear Time Complexity} 
\author{\IEEEauthorblockN{Kaan Gokcesu}, \IEEEauthorblockN{Hakan Gokcesu} }
\maketitle

\begin{abstract}
	We propose a nearest neighbor based clustering algorithm that results in a naturally defined hierarchy of clusters. In contrast to the agglomerative and divisive hierarchical clustering algorithms, our approach is not dependent on the iterative working of the algorithm, in the sense that the partitions of the hierarchical clusters are purely defined in accordance with the input dataset. Our method is a universal hierarchical clustering approach since it can be implemented as bottom up or top down versions, both of which result in the same clustering. We show that for certain types of datasets, our algorithm has near-linear time and space complexity.
\end{abstract}

\section{Introduction}

In the problem of clustering, the aim is to group together the data points in meaningful ways such that similar data points belong to the same groups \cite{jain1999data,popat2014review}. This data analysis is commonly used in many distinct fields of research \cite{ghosal2020short,anderberg2014cluster} including signal processing \cite{ratton1997comparative,mendizabal2018genomic,theis2006median,stankovic2019graph,istepanian2011comparative,gokcesu2021nonparametric}, bioinformatics \cite{zou2020sequence,gokcesu2018adaptive,masood2015clustering,gokcesu2018semg,higham2007spectral}, machine learning \cite{nagamma2015improved,vanli2016sequential,zhang2012traffic}, recommender systems \cite{gentile2014online,neyshabouri2018asymptotically,song2014clustering,gokcesu2018online,bouneffouf2019optimal,nguyen2014dynamic,vural2019minimax}, anomaly detection \cite{munz2007traffic,gokcesu2017online,syarif2012unsupervised,gokcesu2018sequential,agrawal2015survey,delibalta2016online,fu2005similarity}.
In general, clustering is a form of unsupervised classification problem, where the samples are distributed to classes according to a measure of similarity (generally a metric or distance function) \cite{fukunaga2013introduction}. In actuality, the dissimilarity measure need not be a distance function in the mathematical sense and different approaches are also possible \cite{rammal1986ultrametricity}; but we focus on the more popular metric space.

There exist many varying criteria, which can be utilized to define clusters or partitions of a given dataset. However, such definitions are generally subjective and heavily depend on the specific application that needs clustering \cite{jain1999data}. For this reason, it is usually not apparent, which clustering technique creates a better partitioning of the dataset, albeit some considerations can be made in accordance with the application at hand \cite{popat2014review}. 
%The clustering of a dataset depend on the specific method that is utilized to create those clusters. 

All in all, the cluster analysis does not refer to any single algorithm but an approach; where given specific problems, various algorithms can be utilized that differ heavily in their design and understanding of what a cluster means. Traditionally, the notion of cluster is used to mean either of the following: sample groups with small distances in between, sample groups in densely populated regions of the data space, or sample groups that conform to some distribution \cite{anderberg2014cluster}. 
It is generally an optimization problem including multiple objectives, where the selection and design of the clustering algorithms or the appropriate parameters (including the metric) is dependent on the input dataset and what we aim to achieve from this analysis \cite{jain1999data}. Thus, clustering is a knowledge discovery process, where the relevant input and even the dataset itself may need to be processed until a satisfactory analysis is made.

The cluster analysis is a very rich field of research, there exist three main topics in which a clear distinction can be made \cite{jain1999data}.

\begin{enumerate}
	\item If we focus on the algorithm itself
	\begin{itemize}
		\item Parametric clustering: This approach is utilized when some information about the clusters are known a priori. Based on this prior knowledge and corresponding assumptions, some conclusions are drawn \cite{reynolds2009gaussian}.
		\item Nonparametric clustering: In this approach, no information is available a priori. To this end, some local criterion is designed and generally the consequent high density regions are identified in the dataset \cite{fukunaga2013introduction}.
	\end{itemize}

	\item If we focus on the data representation:
	\begin{itemize}
		\item Central clustering: In these techniques, the clusters resulting from the dataset are summarized by their parameters such as the mean and covariance \cite{gersho2012vector}.
		\item Pairwise clustering: In these techniques, the data is represented by a distance matrix, which measures the similarity for each pair of points. The distance function to be used is up to the design \cite{hart2000pattern,hofmann1997pairwise}.
	\end{itemize}
	
	\item If we focus on the clustering solution:
	\begin{itemize}
		\item Hierarchical clustering: These techniques (agglomerative and divisive) produce nested clusters, where all of the clusters are in a hierarchy and can be analyzed accordingly \cite{jain1988algorithms,basalto2007hausdorff,basalto2008hausdorff}.
		\item Partitional clustering: These techniques produce a single unique partitioning of the dataset \cite{celebi2014partitional}.
	\end{itemize}
\end{enumerate}

In this work, we focus on the nonparametric approaches, where instead of selecting a relevant parameter arbitrarily, we aim to create algorithms that tune such parameters inherently. While parametric approaches may be useful in some scenarios, they require a priori knowledge, which may not be present in general. 

Moreover, even though central techniques to create clusters can be meaningful in specific applications, their use case maybe limited in general. In some datasets, the samples may not be distributed properly with a multimodal distribution, and the optimization of centralizing parameters may not be meaningful. To this end, we focus on the pairwise clustering analysis, which is intuitive in the absence of prior information since the dataset is all we are provided with. 

Finally, we are interested in the hierarchical clustering techniques since it is a more detailed cluster analysis of a given data set. Although, the partitive techniques are by no means lacking, the hierarchical techniques have greater potential for knowledge discovery. 

\section{Natural Hierarchical Clustering}\label{sec:hierarchy}

In this section, we propose the notion of natural hierarchical clustering. We start with some important preliminaries.
\subsection{Preliminaries}
Given a multidimensional dataset $\mathcal{X}=\{x_n\}_{n=1}^N$, where $x_n\in\Re^d$ for all $n$ and some dimension $d$. We want to create meaningful clusters $\{\mathcal{C}_i\}_{i=1}^I$ from the dataset $\mathcal{X}$ such that they are disjoint and cover $\mathcal{X}$, i.e.,
\begin{align}
	\mathcal{C}_i\cap\mathcal{C}_j=\emptyset,&&\forall i,j: i\neq j,\\
	\cup_{i=1}^I\mathcal{C}_i=\mathcal{X}.
\end{align}

It depends on the formulation to decide how to create these clusters $\{C_i\}_{i=1}^I$ and even their number $I$. For many applications, the number of clusters may not be known and may not even be easy to estimate. 

One approach is utilizing the dissimilarity of the elements in the dataset, which is measured by a predetermined metric.
\begin{definition}\label{def:d}
	Let us have a metric or distance function $d(\cdot,\cdot)$, which measures the dissimilarity between the data points. By definition, the function $d(\cdot,\cdot)$ satisfies the axioms of 
	%\begin{align*}
	%	d(x,y)=&0 \iff x=y,\\
	%	d(x,y)=&d(y,x),\\
	%	d(x,y)\leq& d(x,z)+d(z,y),
	%\end{align*}
	identity of indiscernibles, symmetry and triangle inequality.
\end{definition}

\begin{remark}
	Given a metric $d(\cdot,\cdot)$ as in \autoref{def:d}, $d(x_i,x_j)$ measures the dissimilarity between $x_i$ and $x_j$ such that greater $d(x_i,x_j)$ means greater dissimilarity.
\end{remark}

One can postulate that more similar points should belong to the same cluster and more dissimilar points should belong to different clusters. In hierarchical clustering, this notion is generalized such that more similar clusters should belong to the same cluster and more dissimilar clusters should belong to different clusters. To this end, there is a need to also define a dissimilarity function for the clusters in general. 
\begin{remark}\label{thm:linkage}
	Some dissimilarity measures (linkage criteria) for two clusters $\mathcal{C}_i$ and $\mathcal{C}_j$ are as follows:
	
	\begin{itemize}
		\item Maximum Distance: $\max_{x\in\mathcal{C}_i}\max_{y\in\mathcal{C}_j} d(x,y)$
		\item Minimum Distance: $\min_{x\in\mathcal{C}_i}\min_{y\in\mathcal{C}_j} d(x,y)$
		\item Average Distance: $\frac{1}{\abs{\mathcal{C}_i}\abs{\mathcal{C}_j}}\sum_{x\in\mathcal{C}_i}\sum_{y\in\mathcal{C}_j}d(x,y)$
	\end{itemize}
\end{remark}

There are more varying dissimilarity (linkage) definitions. All in all, using these cluster dissimilarities, a hierarchy of clusters is built. 

\begin{remark}
	The hierarchical clustering can be divided into two different approaches, which are agglomerative and divisive.
	\begin{enumerate}
		\item Agglomerative clustering is a bottom up approach, where each data point starts by defining its own cluster and pairs of clusters are combined as we move to the top of the hierarchy.
		\item Divisive clustering is a top down approach, where all data points start in a single cluster and clusters are split into pairs as we move to the bottom of the hierarchy.
	\end{enumerate}
\end{remark}

The results of hierarchical clustering are usually presented in a dendrogram \cite{nielsen2016hierarchical}. The combination or division is done according to the minimum or maximum dissimilarity, respectively, between cluster pairs. 

\subsection{A Well Defined Hierarchy of Clusters}
Our goal is to create a well defined hierarchy of clusters. For this reason, we postulate four important notions. The first one is regarding the distinct creation of the agglomerative and divisive approaches. In general, the combinations or divisions are selected in a greedy manner. Because they are selected greedily, they generally do not produce the same hierarchy of clusters when done from top or bottom (agglomerative or divisive), even when the same dissimilarity measure is used. This lack of universality is a problem and we make the following notion for this reason.

\begin{notion}
	A well defined hierarchy of clusters should be defined independently from its method of construction (e.g., iterative approach), i.e., each level in the hierarchy should be defined independently from our levels.
\end{notion}

Moreover, in hierarchical clustering approaches, the clusters are combined (or divided) in a pair by pair manner, which is arbitrarily greedy. In a hierarchy, limiting the number of child nodes of a parent node may prove to be detrimental. These approaches can be more generalized for varying number of clusters' combination (or division) but optimizing such parameters may prove to be much harder. However, this does not change the situation that if need be, more than two clusters should be combined or divided into given the circumstances; which results in the following notion.

\begin{notion}
	In a well defined hierarchy of clusters, each parent cluster can be made up of more than two clusters in accordance with its definitions.
\end{notion}

Moreover, the linkage criteria (such as the ones in \autoref{thm:linkage}) may seem arbitrary and their utilities are heavily dependent on the application since all of them have their own shortcomings. This kind of arbitrariness is undesirable. However, we observe that all of these linkage criteria are purely dependent on the pairwise distance between the data points. Henceforth, we postulate the following.

\begin{notion}
	The decision to whether the clusters should be combined (or divided) should be made in accordance with using purely the pairwise distances between the cluster samples, instead of arbitrary measures.
\end{notion}

Furthermore, in all of these hierarchical approaches, aside from the dendrogram, there does not exist any well defined partitioning of the sample dataset. To decide on a single partition of the clustering is generally dependent on many heuristics. These can be limitations on the number of clusters or their sizes. We also cannot arbitrarily choose any level in the hierarchy since they are not natural results of the datasets but the utilized algorithm, which results in the following notion.

\begin{notion}
	In a well defined hierarchy of clusters, we should be able to determine a partitional clustering that naturally results from the provided dataset.
\end{notion}

Henceforth, the question is to create a clustering algorithm that can answer all of the questions above and satisfy all these desirable properties. In the next section, we show that there exists a scheme that fits to these requirements naturally.

\section{$k$ Nearest Neighbor Clusters}\label{sec:nearest}
In this section, we show that a nearest neighbor based clustering most naturally results in many desirable properties. 

\subsection{Nearest Neighbor Cluster}
We start our design with the following assumption.

\begin{assumption}
	We assume that there are no outliers in the set $\mathcal{X}$ in the sense that for a natural clustering of $\mathcal{X}$, every cluster is of size strictly greater than $1$, i.e., $|\mathcal{C}_i|>1$ for all $i$.
\end{assumption}

We observe that in all of the different agglomerate hierarchical clustering approaches, the cluster combinations start by joining the smallest separated points. We extend this observation and make the
following notion.

\begin{notion}
	Each point should belong to the same cluster as its nearest point. If $\{\mathcal{C}_i\}_{i=1}^I$ is a clustering of $\mathcal{X}$, we have
	\begin{align*}
		\argmin_{y\in\mathcal{X}}d(x,y)\in\mathcal{C}_i,&& \forall x\in\mathcal{C}_i,&&&\forall i.
	\end{align*}
\end{notion} 

This notion clusters the data points in a maximum likelihood sense. We point out that such a notion naturally results in a unique clustering of the dataset.

\begin{remark}
	Although this clustering results in clusters with size at least $2$, it may create a lot of small clusters. Stopping at this point may seem arbitrary and one can also postulate that not only the nearest neighbor, but a point should be in the same cluster as its $k$ nearest neighbors. Hence, 
	\begin{align*}
		\mathcal{N}_k(x)\subset\mathcal{C}_i,&& \forall x\in\mathcal{C}_i,&&&\forall i,
	\end{align*}
	where $\mathcal{N}_k(x)$ is the set of $k$ nearest neighbors of $x\in\mathcal{X}$.
\end{remark}

One question that arises is the choice of $k$. Although $k\geq 1$, the exact value is hard to determine, which makes this approach (as it is) a parametric one.

\begin{definition}\label{def:C_iI_k}
	Let us define a cluster set $\{\mathcal{C}_i^k\}_{i=1}^{I_k}$, which satisfies the following:
	\begin{align*}
		I_k=\argmax_I I:&& \mathcal{C}_i^k=\cup_{x\in\mathcal{C}_i^k}\mathcal{N}_k(x), \forall i\in\{1,\ldots,I\},
	\end{align*}
	i.e., $I_k$ is the maximum number of clusters such that each cluster includes the $k$ nearest neighbors of its every element.
\end{definition}

\begin{proposition}\label{thm:Cikcar}
	For any $k$, we have the following result
	\begin{align*}
		\abs{\mathcal{C}_i^k}\geq k+1, &&\forall i.
	\end{align*}
	\begin{proof}
		Let $x\in \mathcal{C}_i^k$ for some $x\in\mathcal{X}$. Since $\mathcal{N}_k(x)\subset \mathcal{C}_i^k$, we reach the result.
	\end{proof}
\end{proposition}

\begin{lemma}
	Given $k$, the number of clusters $I_k$ and the corresponding set of clusters $\{\mathcal{C}_i^k\}_{i=1}^{I_k}$ are unique.
	\begin{proof}
		Given $k$, we know that each element should be in the same cluster with its $k$ nearest neighbors. Hence, each element is connected with its $k$ nearest neighbors and exist in some cluster all together. If we successively combine these elements, we will end up with the sets $\{\widetilde{\mathcal{{C}}_i^k}\}_{i=1}^{\widetilde{I_k}}$ for some $\widetilde{I_k}$. To satisfy the union condition in \autoref{def:C_iI_k}, the cluster set $\{\mathcal{C}_i^k\}_{i=1}^{I_k}$ can only be made up of the clusters $\{\widetilde{\mathcal{{C}}_i^k}\}_{i=1}^{\widetilde{I_k}}$. However, combining any two clusters will decrease the number of clusters. Hence, we have $\{{\mathcal{{C}}_i^k}\}_{i=1}^{{I_k}}=\{\widetilde{\mathcal{{C}}_i^k}\}_{i=1}^{\widetilde{I_k}}$, which concludes the proof.  	
	\end{proof}
\end{lemma}

\subsection{Existence of a Natural Hierarchy}
We observe that the nearest neighbor clustering implies a natural hierarchical graph. Instead of the seemingly arbitrary hierarchical combination by linkage techniques, where each time only two clusters are greedily combined together, this provides a more fundamental analysis. 

\begin{lemma}\label{thm:CikCjk+1}
	Let us have $\{\mathcal{C}_i^k\}_{i=1}^{I_k}$ and $\{\mathcal{C}_i^{k+1}\}_{i=1}^{I_{k+1}}$ for any $k$. We have
	\begin{align}
		\mathcal{C}_i^{k+1}\equiv \cup_{j\in J_i} \mathcal{C}_j^k,
	\end{align}
	where $J_i$ is a set of indices dependent on $i$ and $\abs{J_i}\geq 1$.
	\begin{proof}
		The proof comes from the fact that the set sequence $\mathcal{N}_k(x)$ is monotone increasing, i.e.,
		\begin{align}
			\mathcal{N}_1(x)\subset \mathcal{N}_2(x)\subset\ldots\subset \mathcal{N}_k(x)\subset\ldots\subset\mathcal{N}_{N-1}(x),
		\end{align}
		$\forall x\in\mathcal{X}$.
		Hence, for any $x\in\mathcal{X}$, if $x\in \mathcal{C}_i^{k+1}$ and $x\in\mathcal{C}_j^{k}$ for some $i,j$; we have 
		\begin{align}
			\mathcal{C}_j^k\subseteq\mathcal{C}_i^{k+1},
		\end{align}
		which implies from \autoref{def:C_iI_k}
		\begin{align}
			\mathcal{C}_i^{k+1}\equiv \cup_{j\in J_i} \mathcal{C}_j^k,
		\end{align}
		for some $J_i$ with cardinality at least $1$ and concludes the proof.
	\end{proof}
\end{lemma}

\begin{corollary}
	For any $k$, we have
	\begin{align}
		I_{k+1}\leq I_k.
	\end{align}
	\begin{proof}
		The result follows from \autoref{thm:CikCjk+1}.
	\end{proof}
\end{corollary}

\begin{theorem}
	The set of clusters $\{\mathcal{C}_i^k\}_{i=1}^{I_k}$ for all $k$ directly implies a hierarchical clustering, where the bottom layer has $I_0=N$ clusters, which are the individual samples, i.e.,
	\begin{align*}
		\mathcal{C}_i^0=\{x_i\},&& i\in\{1,2,\ldots,N\},
	\end{align*}
	and the top layer has $I_{k^*}=1$ (for some $k^*$) cluster, which is the whole set
	\begin{align*}
		\mathcal{C}_1^{k^*}=\mathcal{X}.
	\end{align*}
	\begin{proof}
		The proof follows from the relation in \autoref{thm:CikCjk+1}, where each cluster in a certain level of the hierarchy is composed of its child clusters. At the bottom, we look at the $0$ nearest neighbors, hence, each cluster is a sample itself. At the top, all of the clusters are combined to satisfy the $k^*$ nearest neighbor criterion to create the whole set, where $k^*$ is purely dependent on the dataset. If for all $k$ nearest combinations, we disregard the iterations where the sets do not change, we will end up with a hierarchy of clusters. Each cluster can be traced to its at least two number of child clusters, which concludes the proof.
	\end{proof}
\end{theorem}

\begin{remark}
In general, there are no bounds on the number $k^*$ except the trivial bound of $k^*\leq N/2$, which comes from \autoref{thm:Cikcar}. 
\end{remark}

In the next section, we show that this hierarchy can be constructed in near-linear time in certain datasets.

\section{Near-Linear Complexity Implementation}\label{sec:implement}
To create the hierarchy of clusters, we first need to calculate the distance between the data point pairs in $\mathcal{X}$. 
\subsection{Distance Matrix}
We start by creating the distance matrix $\boldsymbol{D}$.

\begin{definition}
	The distance matrix $\boldsymbol{D}$ contains each pairwise distance of its sample dataset $\mathcal{X}=\{x_n\}_{n=1}^N$, i.e.,
	\begin{align*}
		\boldsymbol{D}(i,j)=d(x_i,x_j), &&\forall i,j\in\{1,2,\ldots,N\}.
	\end{align*}
\end{definition}

The time it takes to calculate the distance function $d(x_i,x_j)$ is dependent on the distance function $d(\cdot,\cdot)$ itself. However, we can analyze the total number of calculations, which is $O(N^2)$.

After calculating the distance matrix $\boldsymbol{D}$, we can order each row in $O(N\log N)$ time. Hence, it takes $\tilde{O}(N^2)$ time.\footnote{$\tilde{O}(\cdot)$ is the soft-O notation, which ignores the logarithms.}

After the creation of the ordered distances, we can start creating the hierarchy.
Starting with each sample in its own cluster, one can combine the nearest neighbors together to create the clusters at the first level of the hierarchy, which takes $O(N)$ time. Continuing to do so by incrementing the nearest neighbor size one by one, we can complete the hierarchy and at the top we will have a single cluster which is the sample set $\mathcal{X}$. If we have $\mathcal{C}_1^{k^*}\equiv\mathcal{X}$ for some $k^*$, the hierarchy will be created in $O(k^* N)$ time.

In general, in the worst case, the hierarchy will be created in $\tilde{O}(N^2)$ number of calculations. However, we observe that, the time complexity is dominated by the calculation of the distance matrix in the first place.

\subsection{Nearest Neighbor Search Methods}
Efficient methods of calculation can be considered. Nearest neighbor search is a well-studied topic by itself \cite{clarkson2006nearest}. There exists many algorithms in literature to expedite this part of the process. Some popular examples are the use of hierarchical data structures such as k-d tree and ball tree \cite{de2008orthogonal,omohundro1989five}. In all such algorithms, the worst case complexity is not much different than the brute force $O(N^2)$. However, for well distributed datasets, they can be efficient. For example, k-d tree has average $O(\log N)$ complexity to find the nearest neighbor of a sample in randomly distributed datasets \cite{chanzy2001analysis}. Moreover, for Euclidean space, it is possible to find $k$ nearest neighbors of every sample in $\widetilde{O}(k^2 N)$ time \cite{ma2019true}.

Thus, using an appropriate nearest neighbor algorithm, it is possible to efficiently calculate the $k$ nearest neighbors of each sample. Since $k^*$ is not known a priori, we can utilize a doubling trick approach to find all $k^*$ nearest neighbors in $\widetilde{O}(p(k^*)N)$ calculation time, where $p(k^*)$ is polynomial in $k^*$. Hence, using such efficient methods, the construction of the whole hierarchy can be concluded in time complexity of $\widetilde{O}(p(k^*) N)$.

\subsection{Connectivity in Random Graphs}
However, the question still remains about how $k^*$ is bounded. First of all, we observe that if for some $k^*$, we end up with a single cluster, this will be true for all $k\geq k^*$. Hence, $k^*$ is actually the minimum number of nearest neighbors that can create a singular cluster. 

Let us create a graph where each vertex is a data point and connect them by the directed edges, where if an edge $n\rightarrow m$ exists, $x_m$ is one of the $k$ nearest neighbors of $x_n$. Hence, each vertex $x_n$ will have $k$ outgoing edges. The problem is for what $k$, this nearest neighbor graph is connected. Fortunately, there exists many promising results. When the data points are nicely distributed, it is proven that $k^*$ is upper bounded by $O(\log N)$, when $N$ goes to infinity \cite{balister2005connectivity}. Although, we cannot make such a claim for general datasets or even general independent identically distributed ones; it still provides promising results, since this $O(\log N)$ connectivity surfaces in many different types of situations \cite{fienberg2012brief,erdHos1960evolution,posa1976hamiltonian}.

Henceforth, when the data points are nicely distributed, the creation of the cluster hierarchy can be done in $\tilde{O}(N)$ time, since $k^*$ is $O(\log N)$. 

\subsection{Choice of a Cluster Partition} 
Even though we may be able to create the cluster hierarchy efficiently in $\widetilde{O}(N)$ time, there still remains the question of which clustering to choose. A hierarchy of clusters is informative as itself and its analysis can be left to the operator depending on the application. However, in some scenarios, it may be desirable to acquire a single clustering for use.

We see that a straightforward choice for the clustering is the coarsest clustering in the hierarchy, which is the level just before the top. However, we encounter an over-combination issue, where some clusters may be combined prematurely because of their underrepresentation. An obvious example for this is a multimodal distribution with close modalities of small probabilities.

Thus, a desired cluster may be further down the hierarchy. However, choosing which cluster to split or preserve is not an easy choice. To this end, we postulate that two clusters can be considered overly combined when they have no choice apart from combining. This issue is apparent from \autoref{thm:Cikcar}, where the cluster cardinalities have a lower bound based on the nearest neighbors. To avoid over-combination, we can proclaim that if a cluster created from $k$ nearest neighbors have a size of $k+1$, this is a bona fide individual cluster, which will be represented as itself in our final choice of clustering. However, in the hierarchy, there may exist multiple clusters that satisfies this, whose intersection is nonempty. To create mutually exclusive clusters, we can choose the cluster that is higher in the hierarchy.

\section{Conclusion}
We proposed a nearest neighbor based clustering algorithm that results in a naturally defined hierarchy of clusters. Our method is a universal hierarchical clustering approach since it can be implemented as bottom up or top down versions, both of which result in the same clustering. We show that for certain datasets, our algorithm has near-linear $\widetilde{O}(N)$ complexity.

\bibliographystyle{IEEEtran}
\bibliography{double_bib}
\end{document}